\documentclass[11pt]{article}
\usepackage[a4paper,hmarginratio=1:1,vmarginratio=2:3,
totalwidth=15.1cm,totalheight=22.5cm]{geometry}
\usepackage{bm,epstopdf,epsfig,amsmath,amssymb,amsfonts,colordvi,latexsym,comment,cancel,
verbatim}
\usepackage{slashed}
\usepackage{graphicx,graphics}
\usepackage[font=md,captionskip=8pt]{subfig}
\usepackage[usenames,dvipsnames]{color}

\usepackage{setspace}

\newcommand{\cL}{{\cal L}}

\newcommand{\cO}{{\cal O}}

\newcommand{\ra}{\rightarrow}
\newcommand{\be}{\begin{equation}}
\newcommand{\ee}{\end{equation}}
\newcommand{\bea}{\begin{eqnarray}}
\newcommand{\eea}{\end{eqnarray}}
\newcommand{\Ra}{\Rightarrow}

\newcommand{\baa}{\begin{array}}
\newcommand{\eaa}{\end{array}}
\long\def\symbolfootnote[#1]#2{\begingroup
\def\thefootnote{\fnsymbol{footnote}}\footnote[#1]{#2}\endgroup}
\newcommand{\laf}{\lambda_\phi}
\newcommand{\lam}{\lambda_m}
\newcommand{\las}{\lambda_\sigma}

\begin{document} 
\begin{flushright}
CERN-PH-TH-2015-178\\
\end{flushright}

\thispagestyle{empty}

\vspace{2.5cm}

\begin{center}
{\Large {\bf Manifestly scale-invariant regularization

\bigskip
and quantum effective operators}}\\

\vspace{1.cm}

{\bf D. M. Ghilencea \symbolfootnote[1]{E-mail:  dumitru.ghilencea@cern.ch}}$^{\,a,b}$

\bigskip

{\small  $^a$ Theory Division, CERN, 1211 Geneva 23, Switzerland}

{\small $^b$ Theoretical Physics Department, National Institute of Physics}

{\small and Nuclear\, Engineering \, (IFIN-HH)\, Bucharest\, 077125, Romania}

\end{center}

\begin{abstract}
\noindent
Scale invariant theories   are often used to address
the hierarchy problem, however the regularization of their quantum corrections 
introduces a dimensionful coupling (dimensional regularization) or scale
(Pauli-Villars, etc) which break this symmetry  {\it explicitly}. 
We show how to avoid this problem  and  study the implications of a manifestly  scale 
invariant  regularization in (classical) scale invariant theories.
We use a dilaton-dependent subtraction  {\it function} $\mu(\sigma)$ which
after {\it spontaneous} breaking of scale symmetry
generates the usual DR subtraction  {\it scale} $\mu(\langle\sigma\rangle)$.
One consequence is that ``evanescent'' interactions  generated by scale invariance
of the action in $d=4-2\epsilon$ (but vanishing in $d=4$),
 give rise to new, {\it finite}  quantum corrections.
We find a (finite)  correction  $\Delta U(\phi,\sigma)$ to the  
one-loop scalar potential for $\phi$ and $\sigma$, beyond the Coleman-Weinberg term.
$\Delta U$ is  due to an evanescent correction ($\propto\epsilon$) to the  field-dependent masses
(of the states in the loop) which  multiplies the pole 
($\propto 1/\epsilon$) of  the momentum integral, to give a finite quantum 
 result.
$\Delta U$  contains  a non-polynomial  operator $\sim\!\phi^6/\sigma^2$ 
of known coefficient and is independent of the subtraction 
{\it dimensionless}  parameter ($z$).
A more  general $\mu(\phi,\sigma)$ is ruled out since, in their classical decoupling limit,
the visible sector (of the  higgs  $\phi$) and  hidden sector (dilaton $\sigma$) 
still interact at the quantum level, thus the  subtraction function depends on the dilaton only. 
The method  is useful in  models where preserving  scale symmetry at quantum level is important.
\end{abstract}

\newpage

\section{Introduction}

There has recently been a renewed interest in the scale invariance symmetry 
to address  the hierarchy or the cosmological constant problems.
Scale symmetry is not a symmetry of the real world since it requires that no 
dimensionful parameters be present in the Lagrangian. One can impose this symmetry 
on the Lagrangian at the classical level, to forbid any mass scales. 
At the quantum level, the  anomalous   breaking of scale symmetry is in general 
expected.
This is because regularization  of the loop corrections breaks this symmetry {\it explicitly}, 
either  by introducing  a {\it dimensionful} coupling as in dimensional  regularization (DR)
or a mass scale  (Pauli-Villars, cutoff regularizations, etc).
Therefore, the presence of a subtraction (or renormalization) scale $\mu$,
breaks explicitly the (classical) scale invariance of the theory
and ruins the symmetry one actually wants to study.
In DR the  scale $\mu$ relates the dimensionless couplings to the dimensionful ones, 
once the theory is continued analytically from $d=4$ to $d=4-2\epsilon$ dimensions. 
For example, the quartic coupling  ($\lambda_\phi$) of 
a  Higgs-like scalar field $\phi$ acquires a mass dimension, since
\bea\label{y1}
\lambda_\phi = \mu^{2\epsilon} \,\big(\lambda_\phi^{(r)}+\sum_n a_n/\epsilon^n\big)
\eea
 where renormalized $\lambda_\phi^{(r)}$ is dimensionless. 
Thus, the  DR scale $\mu$  breaks scale invariance\footnote{
The exact S-matrix is renormalization scale independent. 
But in perturbation theory we truncate the series, so there is always a 
residual renormalization scheme dependence, which must be be minimised.}.

To avoid this problem in theories in which scale-invariance must be preserved 
during regularization, we use  a scheme in which the couplings become  {\it field-dependent},
something  familiar in string theory.
Indeed, one can replace the scale $\mu$  by a {\it function} $\mu(\sigma)$, $\mu\ra \mu(\sigma)$
 \cite{Englert} (also recent \cite{S1}),
where the  field $\sigma$ is  the dilaton\footnote{We also consider
a more general dependence $\mu=\mu(\phi,\sigma)$ 
where $\phi$ is our  scalar (higgs-like) field.}; for example $\mu(\sigma)\!\propto\! \sigma$.
Of course, $\sigma$ must {\it subsequently}  acquire a non-zero (finite) vev, 
otherwise this relation does not make sense due to
 vanishing power ($\epsilon\!\ra\! 0$) in eq.(\ref{y1}). 
One cannot just  replace $\mu$ by the vev of the 
field $\sigma$, since this would simply bring back the original problem.  
One therefore needs a {\it spontaneous} breaking of the  scale symmetry. 
When the (dynamical) field $\sigma$ acquires a vev, scale invariance 
is broken with the  dilaton $\sigma$ as its Goldstone mode.
This can happen in a framework which includes (conformal) gravity in which
the  dilaton vev is related to the Planck scale.  In this paper we shall not include 
gravity, but assume the dilaton acquires a vev spontaneously (fixed {\it e.g.}  by Planck scale
physics) and search  for solutions  $\langle\sigma\rangle\not\!=\!0$.

The goal of this paper is to investigate the quantum implications of 
a manifestly scale-invariant regularization of a theory that is
 classically scale invariant, using a dilaton-dependent subtraction ``scale''.
This is important since scale invariant theories, see e.g.  \cite{W}-\cite{Abel},
do  not seem to be renormalizable  \cite{S2,GM}, in which case
the regularization of the loops   should preserve all initial symmetries
to avoid regularization artefacts \cite{Bardeen}.
This motivated our work, relevant for theories which study scale invariance at 
quantum level.

This paper continues a previous study \cite{S1}, with notable differences and new results 
outlined below. Consider a scale invariant theory of higgs-like  $\phi$ and dilaton $\sigma$
(other fields may be present).  In ``usual'' DR, quartic couplings 
become dimensionless by replacing $\lambda\ra\mu^{2\epsilon} \lambda$, 
see eq.(\ref{y1}) and this changes the scalar potential 
$V(\phi,\sigma)$. For a field-dependent subtraction function $\mu(\sigma)$ 
this change is $V(\phi,\sigma)\!\ra\! \tilde V\equiv \mu^{2\epsilon}(\sigma)\, 
V(\phi,\sigma)$ which is scale invariant in $d\!=\!4-2\epsilon$ (as it should).
 $\tilde V$ acquired  new ``evanescent''
 interactions\footnote{These are defined as interactions absent in
 $d\!=\!4$ ($\epsilon=0$) but generated in $d=4-2\epsilon$ by scale invariance.} 
 due to the field dependence of $\mu(\sigma)$.
This step generates new, {\it finite} corrections at the quantum level.

For example, we obtain a scale-invariant one-loop 
potential $U(\phi,\sigma)$ which contains a  finite 
(quantum) correction $\Delta U(\phi,\sigma)$ beyond  
the ``usual'' Coleman-Weinberg term \cite{Coleman:1973jx,Gildener:1976ih} 
for the higgs $\phi$ and dilaton $\sigma$. $\Delta U$ is a
new  correction overlooked by previous studies \cite{S1}
and at the technical level it
  arises when  the ``evanescent'' correction ($\propto\! \epsilon$) 
 to the field-dependent masses\footnote{These masses  are 
 $\partial^2\tilde V/\partial\alpha\partial\beta$,
  $\alpha,\beta\!=\!\phi,\sigma$ and contain corrections $\propto\!\epsilon$
 from derivatives of $\mu(\sigma)$, see later.} in the loop, 
multiplies the poles $1/\epsilon$ of the loop integrals,
 thus  giving a finite correction! Note that  
$\Delta U$ contains non-polynomial  operators like  $\phi^6/\sigma^2$ 
of  known  coefficient; such new operators generate in turn
polynomial effective operators, when expanded about
$\langle\sigma\rangle\not=0$ ($\langle\sigma\rangle$ can be arranged to be
 much larger than the electroweak vev $\langle\phi\rangle$, see later).

The subtraction function cannot also depend on the higgs field $\phi$ (as in \cite{S1})
since this would bring non-decoupling quantum effects 
of the visible sector ($\phi$)  to the hidden sector ($\sigma$) even in their 
classical decoupling limit.
As a result, the dilaton-only dependent subtraction function must be of the form
 $\mu(\sigma)=z\,\sigma$ where $z$ is an arbitrary dimensionless parameter. 
Unlike total $U(\phi,\sigma)$,  $\Delta U$ is independent of  the subtraction scale 
($z\langle\sigma\rangle$), being finite. Of course {\it physics} must be independent
of the parameter $z$, so we check that our potential respects the 
Callan-Symanzik equation, see \cite{tamarit} for a  discussion.

Assuming the couplings are initially tuned at the classical level  to enforce a hierarchy
$\langle \phi\rangle\!\ll\! \langle\sigma\rangle$, we show the quantum 
correction to the mass of $\phi$, due to $\Delta U$, remains small
without  additional tuning.
With this scale-invariant   regularization and 
spontaneous breaking of this  symmetry
 one  can  address the  hierarchy problem at higher loops.

In  the case of a field-dependent subtraction function
there is no initial subtraction {\it scale}  present in 
the theory, so there is  no dilatation  anomaly. 
Note that it is possible that a theory be  quantum scale invariant and 
the couplings still run with the momentum scale \cite{tamarit,Armillis}.
One first  performs  loop calculations
with a  field-dependent regularization.
After spontaneous breaking of scale symmetry $\langle\sigma\rangle\not=0$,
the  subtraction  scale  and  all masses and vev's of the theory are generated,
proportional to\footnote{In a scale invariant setup,
in the absence of gravity, one can only predict {\it ratios} of fields vev's.} 
$\langle\sigma\rangle$. After regularization and renormalization
one can  eventually decouple the dilaton, by taking  the limit of vanishing couplings
for it, while keeping the masses of the theory fixed \cite{tamarit}\footnote{With scale 
invariance  broken by  gravity,
the dilaton couplings to matter are expected to be very small.}.

After introducing the model (Section \ref{s2}) we present the scale invariant result of the
one-loop potential for a general subtraction function  (Section \ref{s3}); this function is
shown to depend  on the dilaton only (Section \ref{s4});  
the implications for the   mass of $\phi$ are addressed (Section \ref{s5})
and the Callan-Symanzik equation verified (Section~\ref{s6}), followed by  Conclusions.

\section{A generic  model}\label{s2}

\noindent
Consider a Lagrangian with two real scalar fields 
\bea\label{init}
\cL=\frac{1}{2}\,\partial_\mu \phi\, \partial^\mu \phi
+\frac{1}{2}\,\partial_\mu \sigma\, \partial^\mu \sigma
-V(\phi,\sigma)
\eea

\medskip\noindent
The potential in $d=4$  scale invariant theories has the structure 
$V(\phi,\sigma)=\sigma^4\,W(\phi/\sigma)$ and is an homogeneous function, therefore it 
satisfies the relation\footnote{To find this relation, use that
$V(\alpha\phi,\alpha\sigma)=\alpha^4 V(\phi,\sigma)$ (homogeneous);
differentiate wrt $\alpha$ and set $\alpha=1$.}
$\phi\, \partial V/\partial\phi+\sigma\,\partial V/\partial\sigma= 4 V$.
Using this and with the notation 
$x=\phi/\sigma$, the extremum conditions for $V$ ($\partial_\phi V=\partial_\sigma V=0$)
can be written as $W(x)=W^\prime(x)=0$
if we assume that $\langle\sigma\rangle, \langle\phi\rangle\not=0$. 
One of these conditions fixes the ratio
of the fields vev's, while the second implies a relation (tuning) among the couplings.
If  $x_0=\langle\phi\rangle/\langle\sigma\rangle$  is a solution to these two  conditions, 
then $\langle\phi\rangle$  is proportional to $\langle\sigma\rangle$
which means that  a flat direction exists \cite{W} in this theory,
 along the line in the plane $(\phi,\sigma)$ with $\phi/\sigma=x_0$.
Also since $W(x_0)\!=\!0$ on the ground state, then  $V(\langle\phi\rangle,\langle\sigma\rangle)\!=\!0$.
Thus, in scale symmetric theories a vanishing vacuum energy at a given order 
of perturbation theory demands a tuning of the relation among couplings in that order.

An example of a scale invariant potential that we consider below is
\medskip
\bea\label{v}
V=\frac{\lambda_\phi}{4}\phi^4+\frac{\lambda_m}{2}\,\phi^2\sigma^2+\frac{\lambda_\sigma}{4}\,\sigma^4
\eea

\medskip\noindent
where note that the couplings can depend on $\phi/\sigma$, more fields be present, etc. 

In the simple case the couplings are independent of $\phi/\sigma$,  minimizing this $V$ gives
\medskip
\bea\label{vsol}
\langle\phi\rangle \big(\lambda_\phi \langle \phi\rangle^2+\lambda_m\langle\sigma\rangle^2\big)= 0,\qquad
 \qquad 
\langle\sigma\rangle \big(\lambda_m \langle \phi\rangle^2+\lambda_\sigma\langle\sigma\rangle^2\big)= 0
\eea

\medskip\noindent
One can distinguish the following  situations:

\noindent
{\bf Case a):}
 The ground state is $\langle\sigma\rangle=0$, $\langle\phi\rangle=0$ and
both fields are  massless.

\noindent
{\bf Case b):}
A more interesting case that we study in this paper
is that of spontaneous breaking of scale symmetry
when $\langle\sigma\rangle\not=0$. A solution to both equations in (\ref{vsol}) then exists for 
$\langle\sigma\rangle\!<\!\infty$ (finite), then 
 also $\langle\phi\rangle\not=0$, and 
a non-trivial ground state exists
provided that $\lambda_m^2=\lambda_\phi\lambda_\sigma$ and $\lambda_m<0$, 
which we assume to be true in the following. Then
\bea\label{vv}
\frac{\langle\phi\rangle^2}{\langle \sigma\rangle^2}
=-\frac{\lambda_m}{\lambda_\phi},
\quad\Ra\quad
V=\frac{\lambda_\phi}{4}\Big(\,
\phi^2+\frac{\lambda_m}{\lambda_\phi}\,\sigma^2\,\Big)^2,\qquad 
(\lambda_m^2=\lambda_\sigma\lambda_\phi;\,\lambda_m<0)
\eea
Then a spontaneous breaking of the scale symmetry  implies electroweak symmetry breaking
at tree-level, with a vanishing cosmological constant; {it also demands the existence
of a finite (non-zero) scale $\langle\sigma\rangle$ (unknown) in the theory.}
All scales are then generated by  $\langle\sigma\rangle$\footnote{
Values  $\langle s \rangle\!=\!0$ and $\langle s\rangle\!=\!\infty$, with $s=\phi,\sigma$ are excluded,
unless eq.(\ref{vsol}) is implemented in the sense of a {\it limit} (also the
couplings $\lambda_{m,\phi, \sigma}$ can  depend on the ratio of the two fields but
 such case requires assumptions about this dependence, not studied here).  
Thus, although there are no scales in the initial theory, infinite or 
vanishing values for the fields are excluded in spontaneous breaking of
scale symmetry.}.

Further, one shifts the fields
$\phi\ra \phi+\langle\phi\rangle$ and $\sigma\ra \sigma+\langle\sigma\rangle$
and Taylor expands about the ground state. 
The mass eigenstates are 
$\tilde\phi=\phi\cos\alpha+\sigma\sin\alpha$ and 
$\tilde\sigma=-\phi\sin\alpha+\sigma\cos\alpha$ where $\tan^2\alpha\!=\!-\lambda_m/\lambda_\phi\!>\!0$.
A  flat direction exists, so one field (dilaton $\sigma$) is massless
while the second field $\phi$ that  would be the  Higgs boson in a realistic model,
has  a mass 
\bea\label{vv2}
m_{\tilde\phi}^2=2\,\lambda_\phi\big(1-\lambda_m/\lambda_\phi\big)\,\langle\phi\rangle^2
=- 2\lambda_m\,(1-\lambda_m/\lambda_\phi)\,\langle\sigma\rangle^2\,
\eea

\medskip\noindent
Ultimately, scale invariance is expected to be broken by Planck physics, thus
$\sigma$ will acquire a large vev, $\langle\sigma\rangle \sim M_\textsf{Planck}$.
If one would like to implement a hierarchy with  $m_{\tilde \phi}\sim \langle \phi\rangle\sim \cO(100 
\,{\rm GeV})\ll\langle\sigma\rangle$, one should  tune accordingly the couplings
 $\lambda_\sigma\!\ll\! \vert\lambda_m\vert\!\ll\! \lambda_\phi$.
Such hierarchy of couplings is possible \cite{GGR,KA}.
It was observed \cite{GGR} that the shift symmetry of the dilaton enables
 the couplings $\lambda_{m,\sigma}$ to remain ultra-weak under RG evolution.

One would like to know if at the quantum level this tree-level tuning is enough
or additional tuning (beyond that of $\lambda_m$)  is required to maintain this hierarchy and
 $m_{\tilde\phi}$ light. Indeed, at one-loop  dangerous  corrections can emerge, like
 $m_{\tilde\phi}^2\sim \lambda_\phi^2\langle\sigma\rangle^2$ that would require
additional tuning (of  $\lambda_\phi$) and  would re-introduce the hierarchy problem.

\section{Scale invariance of 1-loop potential and effective operators}\label{s3}

To compute the one-loop potential, consider the DR scheme in
$d=4-2\epsilon$.
Then the mass dimensions are $[L]=d$, $[\phi]\!=\![\sigma]\!=(d-2)/2$; the couplings
$\lambda_\phi$, $\lambda_m$, $\lambda_\sigma$ are dimensionful, $[\lambda]\!=[V^{(4)}]=4-d$.
To render the  couplings dimensionless, one uses  the DR scale $\mu$ and replaces
$\lambda\ra \lambda\mu^{4-d}$. The scale $\mu$ breaks the classical scale invariance. 
To avoid this problem and to  preserve this symmetry during regularization
 replace $\mu$ by a field-dependent function (unknown\footnote{This function is assumed
to be non-zero, finite,  continuous, differentiable, and is determined later.}), so
$\mu^{4-d}\rightarrow \mu(\phi,\sigma)^{4-d}$.
Then the actual Lagrangian is 
\medskip
\bea\label{VV}
L=\frac{1}{2}\,\partial_\mu \phi \partial^\mu \phi 
+ \frac{1}{2}\,\partial_\mu \sigma \partial^\mu \sigma 
- \tilde V(\phi,\sigma),\qquad\qquad
\tilde V(\phi,\sigma)\equiv
\mu(\phi,\sigma)^{4-d}\,V(\phi,\sigma)
\eea

\medskip\noindent
$L$ is scale invariant in $d$ dimensions, $[\tilde V]=d$, $[V]=2d -4$.
Denote by $\tilde M^2$ the field-dependent mass matrix  
\bea
 ({\tilde M}^2)_{\alpha\beta}=
\frac{ \partial^2\tilde V(\phi,\sigma)}{\partial\alpha\partial\beta},
\qquad \alpha,\beta=\phi,\sigma.
 \eea
Then the  one-loop potential, that manifestly respects 
scale invariance, is found from\footnote{Formula (\ref{oneloop})
is derived in the usual diagrammatic approach (for $\tilde V$)
and is valid at one-loop (even in non-renormalizable cases,
if no higher derivative operators exist and kinetic terms are canonical).
Beyond one-loop more vacuum ``bubble'' diagrams exist and then 
formula (\ref{oneloop}) receives corrections \cite{two}.}
\medskip
\bea\label{oneloop}
U&=&\tilde V(\phi,\sigma) -
\frac{i}{2}\,
\,\int \frac{d^d p}{(2\pi)^d}
\,{\rm Tr}\ln \big[ p^2-\tilde M^2(\phi,\sigma)+i\varepsilon\big]
\\
&=&
\tilde V(\phi,\sigma)-\frac{1}{2}\,
\frac{1}{(2\pi)^d}\,\,
\Gamma[-d/2]\,{\rm Tr}\big[\pi {\tilde M}^2(\phi,\sigma)\big]^{d/2}
\\
&=&\tilde V(\phi,\sigma)\,
-\frac{1}{64\pi^2}\,
\sum_{s=\phi,\sigma}
 \tilde M^4_s\,\Big[ \frac{2}{4-d}+\ln\kappa - \ln  {\tilde M}^2_s\Big],
\qquad
\kappa\equiv 4\pi e^{3/2-\gamma_E}
\eea
The sum  is over  the 
eigenvalues\footnote{
 For any values of the fields, $\det (M^2)_{\alpha\beta}$ is positive provided that 
$\lambda_m^2\in [3\lambda_\phi\lambda_\sigma\,(3-2\sqrt 2), 3 \lambda_\phi\lambda_\sigma\,(3+2\sqrt 2)]$
and that $\lambda_\phi$, $\lambda_\sigma$, $\lambda_m$ have all the same sign. 
The eigenvalues are positive if $\lambda_\phi$, $\lambda_\sigma$, $\lambda_m$ are 
positive. If $\lambda_m<0$, $\lambda_{\phi}$, $\lambda_\sigma>0$ one eigenvalue is negative.
For $\lambda_m^2$ outside this interval,
restrictions apply to the ratio $\phi^2/\sigma^2$ for which the eigenvalues are 
both positive. Note that even in the Standard Model, 
the Goldstone mode (negative) field dependent squared mass
leads to complex and infrared divergent corrections
and then only the real part of the potential is included.   
A resummation of higher orders in  $V$ fixes this well known 
problem \cite{espinosa,martin}  (see also \cite{Y,complexV}). Here we proceed in general
and do not study this issue that affects the Coleman-Weinberg term only,
  but refer the reader to \cite{espinosa,martin}.} 
$\tilde M_s^2$ of the matrix  $(\tilde M^2)_{\alpha\beta}$. 
Up to $\cO[(4-d)^2]$ terms
\medskip
\bea\label{m2}
(\tilde M^2)_{\alpha\beta}
&=&
\mu^{4-d}\,\Big[ (M^2)_{\alpha\beta} + (4-d)\,\mu^{-2}\,N_{\alpha\beta}\Big],
\qquad
\alpha,\beta=\{\phi,\sigma\}.
\eea
where
\bea\label{m3}
(M^2)_{\alpha\beta}=V_{\alpha\beta},\qquad
N_{\alpha\beta}&\equiv &\mu\, (\mu_\alpha\, V_\beta +\mu_\beta\, V_\alpha)
 +(\mu\, \mu_{\alpha\beta}  -\mu_\alpha\,\mu_\beta)\,V,
\eea

\medskip\noindent
and  $\mu_\alpha={\partial\mu}/{\partial\alpha}$, $\mu_{\alpha\beta}
={\partial^2 \mu}/{\partial\alpha\partial\beta}$, $V_\alpha=\partial V/\partial\alpha$, 
$V_{\alpha\beta}=\partial^2 V/\partial\alpha\partial\beta$,
  are nonzero field dependent quantities.
From the last two equations one finds, up to $\cO[(4-d)^2]$ terms
\medskip
\bea\label{m4}
\sum_{s=\phi,\sigma}\tilde M_s^4
=\mu^{2\,(4-d)}\,\Big[{\rm Tr}\, M^4 +  2 (4-d)\,\mu^{-2}\,  {\rm Tr}\,(M^2 N)\Big],
\eea
%
Then
\bea
U 
=\mu(\phi,\sigma)^{4-d}
\Big\{
V- \frac{1}{64 \pi^2}\Big[
 \sum_{s=\phi,\sigma} 
M^4_s \,  \Big(  \frac{2}{4-d}- \ln\frac{M_s^2}{\kappa\,\mu^2(\phi,\sigma)} \Big)
+
\frac{4 \,{\rm Tr}\,(M^2\,N)}{\mu^2(\phi,\sigma)}\Big]\Big\}
\eea

\medskip\noindent
The last term  is due to the field dependence
of $\mu$  and its origin  is in the second ``evanescent'' term in the rhs of 
 eqs.(\ref{m2}) which cancels the pole to  give a finite contribution.
We adopt the usual $\overline{MS}$ scheme here, in which case the counterterms 
are\footnote{One can use other subtraction schemes e.g.
$\delta U_{c.t.}\!\!\!=\!\mu^{4-d}
\big[
a_1\phi^4 (1/\tilde\epsilon+c_1)\!+\!
a_2\phi^2\sigma^2 (1/\tilde\epsilon+c_2)\!+\!
a_3\sigma^4 (1/\tilde\epsilon+c_3)\big]$
where we denoted $1/\tilde\epsilon\equiv 2/(4-d)+\ln\kappa-3/2$.
The case of $\overline{MS}$ corresponds to $c_1=c_2=c_3=0$.}
\medskip
\bea\label{ct}
\delta U_{\rm ct}
=\frac{\mu(\phi,\sigma)^{4-d}}{64\pi^2}\sum_{s=\phi,\sigma}
M^4_s \,\Big(\frac{2}{4-d}+\ln\kappa-\frac{3}{2} \Big),
\eea

\medskip\noindent
where $\sum_{s=\phi,\sigma} M_s^4=V_{\phi\phi}^2+V_{\sigma\sigma}^2+2V_{\phi\sigma}^2$.
Using  
\bea\label{Z}
\delta U_{c.t.}\equiv \mu^{4-d} 
\big[ 1/4\, (Z_{\lambda_\phi}-1) \lambda_\phi\,\phi^4+
 1/2\,(Z_{\lambda_m}-1)\lambda_m\,\phi^2\,\sigma^2
+1/4 \,(\lambda_\sigma-1)\,\lambda_\sigma\,\sigma^4\big]
\eea one finds
the renormalization coefficients
\bea
Z_{\lambda_\phi}&=& 
1+\frac{1}{8 \pi^2\,(4-d)} (9\lambda_\phi+\lambda_m^2/\lambda_\phi)
\nonumber\\
Z_{\lambda_m}&=& 
1+\frac{1}{8 \pi^2\,(4-d)} (3\lambda_\phi+3\lambda_\sigma+4\lambda_m)
\nonumber\\
Z_{\lambda_\sigma}&=& 
1+\frac{1}{8 \pi^2\,(4-d)} (9\lambda_\sigma+\lambda_m^2/\lambda_\sigma)
\eea

\medskip\noindent
These $Z$'s have  expressions identical to those obtained
at one-loop with $\mu$ a constant.

After adding the counterterms $\delta U_{c.t}$  we can safely take the limit $d\ra 4$ in 
the remaining terms ($\mu\not=0$),  so the renormalized one-loop potential is
\medskip
\bea\label{U}
U(\phi,\sigma)&=&V(\phi,\sigma)+\frac{1}{64\pi^2}\,
\Big\{
\sum_{s=\phi,\sigma} M^4_s(\phi,\sigma)\, \Big(\ln \frac{M^2_s(\phi,\sigma)}{\mu^2(\phi,\sigma)}
-\frac{3}{2}\,\Big)
+\Delta U(\phi,\sigma)
\Big\}
\nonumber\\[4pt]
\Delta U
\!\!&=&\!\! \frac{-4}{\mu^2} \Big\{
V\,\big[ (\mu\mu_{\phi\phi}-\mu_\phi^2)\,V_{\phi\phi}
+2\,(\mu\mu_{\phi\sigma}-\mu_\phi\mu_\sigma) V_{\phi\sigma}
+(\mu\mu_{\sigma\sigma}-\mu_\sigma^2)\,V_{\sigma\sigma}\big]
\nonumber\\[4pt]
&+&
2\mu\,(\mu_\phi\,V_{\phi\phi}+\mu_\sigma\,V_{\phi\sigma})\,V_\phi
+2\mu\,(\mu_\phi\,V_{\phi\sigma}+\mu_\sigma\,V_{\sigma\sigma})\,V_\sigma\Big\}
\eea

\medskip\noindent
In the above  $M_s^2$ ($s=\phi,\sigma$) 
  are the eigenvalues of the matrix $V_{\alpha\beta}$,
given by the roots of equation  $\rho^2-\rho\,(V_{\phi\phi}+V_{\sigma\sigma})+
(V_{\phi\phi} V_{\sigma\sigma}-V_{\phi\sigma}^2)=0$ \footnote{
For the particular $V$ of eq.(\ref{v}), the eigenvalues
 $M_s$ ($s=\phi,\sigma$) of $(M^2)_{\alpha\beta}$ are
\bea
\label{ms}
 M^2_s&=& (1/2)\big[ \,\nu \pm  \sqrt \Delta\big],\qquad\nu\equiv 
(3\lambda_\phi +\lambda_m)\phi^2+(3\lambda_\sigma+\lambda_m)\sigma^2,
\\
\Delta &=&
(3\lambda_\phi-\lambda_m)^2\,\phi^4 +(3\lambda_\sigma-\lambda_m)^2\,\sigma^4
+2\phi^2\sigma^2\,[ 3\lambda_m (\lambda_\phi+\lambda_\sigma)
-9\lambda_\phi\lambda_\sigma+7\lambda_m^2
\big]\nonumber
\eea
\vspace{-0.2cm}}.

Eq.(\ref{U})  is a  scale-invariant one-loop result. It is a 
modified version of the  Coleman-Weinberg potential 
(recovered if  $\mu$ is a  constant)  and contains an
additional correction ($\Delta U$).
Note that $\Delta U$ is not exactly  a counterterm 
but  a {\it finite} one-loop effect induced by scale invariance. 
It is generated when the ``evanescent''
coefficient ($4-d$) in the field-dependent masses of eq.(\ref{m2}),
 multiplies the pole $1/(4-d)$ of the one-loop integral\footnote{In higher
orders, a $n$-loop pole $1/(4-d)^n$, upon multiplication by the  $4-d$ coefficient 
 will actually generate   a $1/(4-d)^{n-1}$ pole, i.e. what we consider usually to
account for $n-1$ loop effects. Thus,  the order of the singularity is not
identical to the loop order  in this case.}.
This effect is missed in calculations that 
are not  scale invariant such as the usual DR scheme.
Note  also that $\Delta U$  vanishes on the tree-level ground state.

 $\Delta U$  contains  non-polynomial operators. Even in the  minimal case of taking
$\mu\sim \sigma$, then  the terms in $\Delta U$  proportional to $V V_{\sigma\sigma}$ 
contain a $\phi^6/\sigma^2$ term.   
Similar  effective operators are expected to be generated in higher orders.
Further, one can Taylor expand the expression of the potential about the ground state, using
$\sigma=\langle\sigma\rangle+ \delta\sigma$, with $\delta \sigma$ a quantum fluctuation.
 When doing so, the  operator $\phi^6/\sigma^2$
 becomes  a series of effective (polynomial) operators
\bea
\frac{\phi^6}{\sigma^2}=\frac{\phi^6}{\langle\sigma\rangle^2}
\Big(
1- \frac{2\delta\sigma}{\langle\sigma\rangle}
+\frac{3\delta\sigma^2}{\langle\sigma\rangle^2}+\cdots\Big).
\eea

To proceed further, one needs the general expression of the function  $\mu=\mu(\phi,\sigma)$. 
Let us first take $\mu=\mu(\sigma)$ only, which will be justified in the next section;
  in this case the only possibility is
\bea\label{miu}
\mu(\sigma)=z\,\sigma
\eea

\medskip\noindent
which, as a ``DR scale'',  requires $\langle\sigma\rangle\not=0$, $\langle\sigma\rangle\!<\!\infty$.
 To be exact, we actually take 
$\mu(\sigma)=z\,\sigma^{2/(d-2)}$ \cite{tamarit}, which accounts for the 
mass dimension of the field $\sigma$. 
For one-loop case only (as here)  it is safe to use at this stage
 its limit for $d\ra 4$, so  $\mu=z\,\sigma$.
Here $z$ is an arbitrary {\it dimensionless} parameter and the dependence of $U$ on $z$
is  equivalent to  the familiar  subtraction {\it scale}  dependence  of $U$ in 
the ``usual'' regularization.
With eq.(\ref{miu}),  one obtains the following form of  $\Delta U$, which is 
independent of $z$
\medskip
\bea\label{delta1}
\Delta U=-\frac{4}{\sigma^2}\,
\big[V_{\sigma\sigma}\,(2 \sigma \,V_\sigma- V)  + 2 \sigma\,V_\phi\,V_{\phi\sigma}\big]
\eea

\medskip\noindent
and only the Coleman-Weinberg term depends on $z$. 
With  $V$ of eq.(\ref{v})  
\medskip
\be\label{tr}
\Delta U =
\frac{\laf\lam \phi^6}{\sigma^2}
-
\big(16 \laf\lam+6\lam^2-3 \laf\las\big)\phi^4
-\big(16\lam+25\las\big)\,\lambda_m\,\phi^2\sigma^2
- 21\las^2\sigma^4
\ee

\medskip\noindent
As anticipated, notice  the presence of the non-polynomial operator $\sim \phi^6/\sigma^2$.
This operator is suppressed at large $\langle \sigma\rangle$ or for small mixing ($\lambda_m$) 
between $\phi$ and the dilaton $\sigma$.
The sign of this operator  is controlled by $\lambda_m$, assuming  
$\lambda_\phi>0$. When $\lambda_m<0$, the term $\lambda_m\phi^6$ 
destabilizes the potential for large values of 
$\phi$. A tuning  $\vert \lambda_m\vert\ll\lambda_\phi$ can 
compensate to  render this term of similar size to $\phi^4$ terms; also
higher loop orders can generate similar effective operators that
may stabilize  the  potential globally.

For the special case of a non-trivial classical vacuum of  eq.(\ref{vv}), 
when  $\lambda_m^2=\lambda_\phi\,\lambda_\sigma$,  eq.(\ref{tr}) becomes
\medskip
\bea\label{delta2}
\Delta U=\frac{\lambda_m}{\lambda_\phi^2\,\sigma^2}\,
\big(\lambda_\phi\,\phi^2+\lambda_m\,\sigma^2\big)
\big[\,\lambda_\phi^2\,\phi^4-4\,\lambda_\phi\,
(4\,\lambda_\phi+\lambda_m)\,\phi^2\,\sigma^2 -21 \lambda_m^2\,\sigma^4\big]
\eea

\medskip\noindent
This expression  vanishes on the tree-level
ground state\footnote{At the loop level the ground 
 state  is changed slightly, but we ignore that effect here.},
(see eq.(\ref{vv}); $\lambda_m\!<\!0$, $\lambda_\phi\!>\!0$).

In conclusion, the  expression of $U$ at one-loop is manifestly scale invariant
\bea\label{fi}
U(\phi,\sigma)=
V(\phi,\sigma)
+\frac{1}{64\pi^2}\,
\Big[
\sum_{s=\phi,\sigma} M^4_s(\phi,\sigma)\, \Big(\ln \frac{M^2_s(\phi,\sigma)}{z^2 \sigma^2}
-\frac{3}{2}\,\Big)
+\Delta U(\phi,\sigma)\Big]
\eea

\medskip\noindent
with $\Delta U$ as in eqs.(\ref{tr}) or (\ref{delta2})
 and  $V$ of eq.(\ref{v}); note that 
the ``standard'' Coleman-Weinberg term is modified
into a  scale invariant form. This is the
main result of this section, valid under our assumption $\mu(\sigma)=z\sigma$. 
Further, $U$  can be Taylor expanded about $\langle\sigma\rangle$.
With no mass scale in the theory, from minimising $U$ 
one can only predict ratios of vev's, so all masses are generated by $\langle\sigma\rangle$
after spontaneous breaking of scale symmetry.

\section{More general $\mu(\phi,\sigma)$ and implications}\label{s4}

The subtraction function could in principle be more general
and could depend on $\phi$ too, $\mu=\mu(\phi,\sigma)$. 
In this section we show that  such dependence is not physical
and  conclude that  $\mu$ must be a function of the dilaton only.
 First,  consider the following example
\medskip
\bea\label{muS}
\mu(\phi,\sigma)=z\, \big(\xi_\phi\,\phi^2+\xi_\sigma\,\sigma^2\big)^{1/2}
\eea

\medskip\noindent
This was used in earlier similar studies \cite{S1,S3} where scale invariant models had
a  non-minimal coupling to gravity,
 with this expression to fix the Planck scale upon spontaneous breaking of scale 
symmetry\footnote{The non-minimal coupling is 
$\cL_G=-\frac{1}{2}\,(\xi_\phi\,\phi^2+\xi_\sigma\,\sigma^2)\,R,$
and is added in some models to generate the Planck mass from $\mu(\sigma,\phi...)$,
in (spontaneously broken) scale invariant theories  \cite{S1,S3}.
The relative signs of $\xi_\phi$, $\xi_\sigma$
are important to ensure a positive Newton constant (for a review see \cite{Oda}).
When going to the Einstein frame, this coupling generates a suppression of the 
tree level potential  by a factor  $1/(\xi_\phi\,\phi^2+\xi_\sigma\,\sigma^2)^2$, while in the
case discussed in the text (where no such coupling is included), such suppression
is shown to be generated in $\Delta U$ at 1-loop, see later.}.
With this $\mu$ and  $V$ of eq.(\ref{v})  one finds that $\Delta U$ contains
 leading  power terms $\phi^8$ and $\sigma^8$ as  shown in
\medskip
\bea
\Delta U=
- (\xi_\phi\phi^2+\xi_\sigma\sigma^2)^{-2}
\Big[
(21\,\lambda_\phi\, \xi_\phi+\lambda_m\,\xi_\sigma)\,\xi_\phi \lambda_\phi \,\phi^8
\!+\!
(21 \lambda_\sigma\, \xi_\sigma+\lambda_m\,\xi_\phi)\,\xi_\sigma \lambda_\sigma\,\sigma^8
\!+\!\cdots\Big]
\eea

\medskip\noindent
The dots stand for remaining $\phi^6\sigma^2$, $\phi^4\sigma^4$ and $\phi^2\sigma^6$ terms
which we do not display since their coefficients are too long.
The coefficients of $\phi^8$, $\sigma^8$ are positive irrespective 
of the values of $\xi_{\phi, \sigma}$, if  $\lambda_m^2\geq 21^2 \lambda_\phi\,\lambda_\sigma$, 
(with $\lambda_{\phi,\sigma}>0$). This condition is not respected on the ground state
of  $V$ (with $\lambda_m^2=\lambda_\phi\lambda_\sigma$).
We thus encounter terms  unbounded from below, that otherwise vanish on the tree level 
ground state.  A small fluctuation about the critical point can then destabilize the potential.

It is intriguing that  even if  the classical  $V$ contains  no interaction terms
between ``visible'' ($\phi$) and ``hidden'' ($\sigma$) sectors, i.e.  $\lambda_m=0$,
such terms are  still  generated by quantum corrections, 
for $\mu(\phi,\sigma)$ of eq.(\ref{muS}).  Indeed, one has
\medskip
\bea
\Delta U\Big\vert_{\lambda_m=0}&=&  
 -3 \,
\Big[\xi_\phi\xi_\sigma\,\big[\lambda_\phi\, (9\lambda_\phi+\lambda_\sigma)\,\phi^6\sigma^2
+\lambda_\sigma (\lambda_\phi+9\lambda_\sigma)\,\phi^2\sigma^6 \big]
\nonumber\\[5pt]
&&\,\, +\,\, 7\,\big(\lambda_\phi^2 \,\xi_\phi^2\,\phi^8
+\lambda_\sigma^2\,\xi_\sigma^2\,\sigma^8\big)\,
-
 (\xi_\phi^2+\xi_\sigma^2)
 \lambda_\phi\lambda_\sigma\,\phi^4\sigma^4 \Big](\xi_\phi\phi^2+\xi_\sigma\sigma^2)^{-2}
\eea

\bigskip\noindent
This simplifies  further if also  $\lambda_\sigma\!=\!\lambda_m^2/\lambda_\phi\!\ra\! 0$, but 
the term  $\propto\xi_\phi \xi_\sigma \lambda_\phi^2\phi^6\sigma^2$  does not vanish.
Such term ultimately arise from the expression of the $\mu$-dependent factor in  $\tilde V$,
via terms like $\mu_\phi\,V_\phi\,V_{\phi\phi}$ and 
$(\mu\,\mu_{\phi\phi}-\mu_\phi^2) V_{\phi\phi}$  in eq.(\ref{U}). 
The two sectors  still ``communicate''   at the quantum level, due 
to scale invariance even if they are classically decoupled!
 This concerning effect is only  removed  for vanishing $\xi_\phi$ or $\xi_\sigma$,
which means  $\mu\propto \sigma$. \footnote{up to a relabeling, see the symmetry
$\phi\leftrightarrow \sigma$, at which stage one decides which field 
denotes the dilaton.}

More generally,  consider 
\medskip
\bea
\mu(\phi,\sigma)=z \,\sigma \,e^{g(\phi/\sigma)}
\eea

\medskip\noindent
Here  $g$ is some arbitrary function of the ratio  $\phi/\sigma$. 
In this case, in  the classical ``decoupling'' limit $\lambda_m\ra 0$, also with 
$\lambda_\sigma=\lambda_m^2/\lambda_\phi\ra 0$, 
there are non-vanishing quantum interactions terms
\bea\label{vqe}
\Delta U\Big\vert_{\lambda_m= 0}=-3 \lambda_\phi^2\,\Big[\, 8 \frac{\phi^5}{\sigma}\,g^\prime(\phi/\sigma)+
\frac{\phi^6}{\sigma^2}\, g''(\phi/\sigma)\Big]
\eea

\medskip\noindent
Again, the two  sectors still communicate at the quantum level only. To avoid  such concerning behaviour, 
we must take  $g\!=\!0$ (or constant)\footnote{We disregard a second solution for which the rhs 
of eq.(\ref{vqe}) vanishes, since it is not continuous in $\phi=0$.}
. Therefore,  the subtraction function is
independent of $\phi$ and  thus $\mu(\sigma)\!=\! z\,\sigma$.  This result  is 
 the minimal scenario  used in the previous  section
and justifies our choice in eq.(\ref{miu}) and our result in eq.(\ref{fi}).
We conclude  that it is the  dilaton  alone that  generates the subtraction scale
after spontaneous breaking of scale symmetry.

\section{The  mass spectrum}\label{s5}

Let us minimise the one-loop potential $U$. We restrict the analysis
to the simpler case of a hierarchy of the couplings
considered in \cite{S1,GGR}. We  take
\medskip
\bea
\lambda_\sigma\ll\vert\lambda_m\vert\ll \lambda_\phi.
\eea

\medskip\noindent
To enforce this hierarchy, introduce $\lambda_m\!=\!\tilde\lambda_m\varepsilon$ and
$\lambda_\sigma\!=\!\tilde\lambda_\sigma\,\varepsilon^2$, where $\varepsilon\!\sim \! 1/M^2_\textsf{Planck}
\ll 1$, and
$\lambda_\phi$, $\tilde\lambda_m$ and $\tilde\lambda_\sigma$ are now of similar magnitude.
One then expands $U$  up to $\cO(\varepsilon^3)\!\sim\! \cO(\lambda_m^3)$
\medskip
\bea\label{rr}
U&=&\frac{\lambda_\phi}{4}\phi^4
+\frac{\lambda_m}{2}\,\phi^2\,\sigma^2
+\frac{\lambda_\sigma}{4}\,\sigma^4
+\,
\frac{1}{64\pi^2}\,\Big\{
M_1^4\,\Big[\ln\frac{M_1^2}{z^2\sigma^2}-\frac{3}{2}\Big]
+M_2^4\,\Big[\ln\frac{M_2^2}{z^2\sigma^2}-\frac{3}{2}\Big]
\nonumber\\[5pt]
&+&
\lambda_\phi\lambda_m\frac{\phi^6}{\sigma^2}
-\big(16\,\lambda_\phi\lambda_m+6\lambda_m^2-3\lambda_\phi\lambda_\sigma\big)
\phi^4
-16\,\lambda_m^2\,\phi^2\sigma^2\Big\}+\cO(\lambda_m^3)
\eea

\medskip\noindent
One  can  minimise $U$ and find the solution for 
$\langle\phi\rangle/\langle\sigma\rangle$
that satisfies $U_\phi=U_\sigma=0$;  $U$ being manifestly scale invariant, 
these conditions ensure a flat direction exists and also that vacuum energy vanishes in this order.
To the lowest order in $\varepsilon$, 
one finds
\bea
\frac{\langle\phi\rangle^2}{\langle\sigma\rangle^2}
=-\frac{\lambda_m}{\lambda_\phi}\,\Big[1- \frac{6 \lambda_\phi}{64\pi^2}\,\big(4\ln3\lambda_\phi
-17/3\big)\Big]+\cO(\lambda^2_m)
\label{ro}
\eea

\medskip\noindent
This brings a  correction to the tree level case, eq.(\ref{vv});
here $\lambda_m<0$, $\lambda_\phi>0$ and $\lambda_m^2=\lambda_\phi\lambda_\sigma$.
To  obtain  eq.(\ref{ro}) we fixed the subtraction parameter $z$  under 
the log term in $U$ to
\medskip
\bea
z=\langle\phi\rangle/\langle\sigma\rangle,\qquad \textrm{then} \qquad
\mu(\langle\sigma\rangle)= \langle\phi\rangle
\eea

\medskip\noindent
on  the ground state. This value for $\mu$ is the  standard choice for the subtraction 
scale,  made to minimize the Coleman-Weinberg log-term dependence on 
it. As mentioned,   $\Delta U$ itself is  scheme-independent 
(being independent of $z$).

The potential in (\ref{rr}) is scale invariant, the 
dilaton remains  massless at one-loop while  the higgs-like 
scalar $\phi$ has a mass
\medskip
\bea
m_{\tilde \phi}^2= \big[U_{\phi\phi}+U_{\sigma\sigma}\big]_{\textsf{min}}
\eea

\medskip\noindent
Let us consider only the contribution $\delta m_{\tilde\phi}^2$ 
from $\Delta U$ alone to the mass of $\phi$.
The interest is to examine if
potentially ``dangerous'' corrections of the type $\lambda_\phi^2 \langle\sigma\rangle^2$, etc,
 can emerge
from the new contribution $\Delta U$.
These would require an additional tuning (of $\lambda_\phi$) beyond that of $\lambda_m$ 
 done at the tree level, in order
 to keep $\phi$ light compared to $\langle\sigma\rangle\sim M_\textsf{Planck}$.
In general, one has
\medskip
\bea\label{deltam}
\delta m_{\tilde \phi}^2
\!\!\!\!&=&\frac{1}{64\pi^2}
\big(\Delta U_{\phi\phi}+\Delta U_{\sigma\sigma}\big)_{\textsf{min}}
\\[5pt]
&=&
\!\!\!
\frac{-\langle\sigma\rangle^2}{32\pi^2} \Big[ 4 \lambda_m^2 (4+13 \rho)
+18 \lambda_\sigma \,(7 \lambda_\sigma-\lambda_\phi\rho)
+\lambda_m \big[25 \lambda_\sigma (1\!+\!\rho)
\!-\!3\lambda_\phi\rho (-32\!+5\rho\!+\!\rho^2)\big]\Big]
\nonumber
\eea

\medskip\noindent
where $\rho=\langle\phi\rangle^2/\langle\sigma\rangle^2$.
This mass correction contains terms  proportional 
to $\lambda_m$ or $\lambda_\sigma=\lambda_m^2/\lambda_\phi\ll \lambda_m$ 
but not to $\lambda_\phi$ alone.
Therefore  no extra tuning  is needed 
beyond that at classical level of  eqs.(\ref{vv}), (\ref{vv2}),
in order to maintain $\delta m_{\tilde\phi}^2$  and 
$m_{\tilde\phi}^2\sim \langle\phi\rangle^2\sim \lambda_m\langle\sigma\rangle^2$
 close to the  electroweak scale.  It is possible that this nice 
behaviour survives to higher or all orders, as a result of the 
manifest scale invariance  and spontaneous breaking of this symmetry.
This suggests that the hierarchy problem could be solved 
with only one  initial (classical) tuning of  $\lambda_m$ (no tuning of higgs self-coupling 
$\lambda_\phi$).

\section{Further remarks}\label{s6}

The method we used to generate dynamically the subtraction scale of the 
 DR scheme as  the dilaton vev deserves further study. 

First, note that the potential $U$ must respect the Callan-Symanzik equation 
i.e. it must be independent of the choice of the  
 dimensionless parameter $z$  and thus of the 
subtraction scale $z\langle\sigma\rangle$  after spontaneous scale symmetry 
breaking \cite{tamarit}. In our
 one-loop approximation this demands that
\medskip
\bea\label{CZ}
\frac{d U}{d\ln z}=\Big
(\frac{\partial U}{\partial \ln z}+\beta_{\lambda_j}\frac{\partial}{\partial \lambda_j}
\Big) U=O(\lambda^3)
\eea

\medskip\noindent
where $U$ is that of eq.(\ref{fi}) and $\Delta U$ of (\ref{tr})
and the Coleman-Weinberg term is the only one that depends {\it explicitly} on $z$.
To check if condition (\ref{CZ}) is respected,  
we  need the one-loop beta functions of the theory; these are obtained 
from the condition that the ``bare'' couplings of the Lagrangian 
are independent of subtraction scale $z\langle\sigma\rangle$,  
where $z$ is arbitrary:
$d ( \lambda_j\,Z_{\lambda_j})/d\ln z=0$, 
where $j=\phi,m,\sigma$ (fixed) and $Z_{\lambda_j}$ are given in eq.(\ref{Z})\footnote{To 
be exact, in $d=4-2\epsilon$, one actually
imposes $d \big((z\sigma)^{2\epsilon} \lambda_j\,Z_{\lambda_j}\big)/d\ln z=0$, giving  that beta functions
are shifted from those above, $\beta_{\lambda_j}=-2\epsilon\lambda_j+(...)$ where $(...)$ denotes 
 the rhs in each of eqs.(\ref{beta}).}.
One finds
\medskip
\bea\label{beta}
\beta_{\lambda_\phi}&=&\frac{d\lambda_\phi}{d\ln z}
=\frac{1}{8\pi^2} \,(9\lambda_\phi^2+\lambda_m^2)
\nonumber\\
\beta_{\lambda_m}&=&\frac{d\lambda_m}{d\ln z}
= \frac{1}{8\pi^2}\, (3\lambda_\phi+4\lambda_m+3\lambda_\sigma)\,\lambda_m
\nonumber\\
\beta_{\lambda_\sigma}&=&  \frac{d\lambda_\sigma}{d\ln z}
=
\frac{1}{8\pi^2}\, (\lambda_m^2+9\lambda_\sigma^2)
\eea

\medskip\noindent
which are the same as in the case the theory was regularized with $\mu=$constant\footnote{This is
expected since we only found new finite terms, but no new counterterms.}.
Using these beta functions one easily checks  that eq.(\ref{CZ}) is respected.
This shows that the change of  parameter $z$ 
 is ``moved'' into the running couplings\footnote{
Thus there  is no dilatation anomaly, yet the couplings still run  as usual
  \cite{tamarit,Armillis}} of the potential
and {\it physics} is indeed independent of $z$:
$U(\lambda_j(z),z)=U(\lambda_j(z_0), z_0)$, where $j=\phi, m, \sigma$ and
$z, z_0$ are different subtraction parameters
(ultimately corresponding to subtraction scales 
$z\langle\sigma\rangle$, $z_0\langle\sigma\rangle$).

Regarding renormalizability of scale invariant models, 
previous studies \cite{S2} identified
 at three-loop order  a UV  counterterm to the original Lagrangian $L$, of the form 
\medskip
\bea
\frac{1}{(16\pi^2)^{3}} \frac{1}{(4-d)^{2}}\Big(\frac{\xi_\phi}{\xi_\sigma\sigma^2}\Big)^2
 (\Box\phi^2)^2
\eea

\medskip\noindent
 In \cite{S2}  $\mu\sim (\xi_\phi \phi^2+\xi_\sigma\sigma^2)^{1/2}$, just like
in eq.(\ref{muS}). This UV divergence was due to a new vertex generated 
by the Taylor expansion of   $\mu(\phi,\sigma)$ 
wrt $\phi$; this vertex is ultimately  due  to new interactions 
 that $\mu(\phi,\sigma)$ itself brought in $\tilde V$ but absent in initial $V$! 
 Given this counterterm,  the theory is  then non-renormalizable and non-local. 
The same conclusion is expected for any subtraction 
function that depends on additional fields other than dilaton.

However, we showed that $\mu\!=z \,\sigma$ (Section~\ref{s4}), so
the above three-loop  counterterm  is absent  because we  have $\xi_\phi=0$.
Despite this,  the standard expectation is that 
higher loop orders still generate  higher dimensional counterterms and the 
theory is  non-renormalizable, due to the presence in $U$ of 
the non-polynomial term $\phi^6/\sigma^2$
(one can still explore the possibility that in a scale symmetry-preserving calculation, 
 all poles in quantum corrected $L$ be those that renormalize 
its initial couplings and fields only (i.e. renormalizability), without other UV
 counterterms. This problem deserves  careful investigation and is beyond the 
goal of this paper).

As a result of a manifestly scale-invariant regularization, 
the (mass)$^2$ of $\phi$ contains:
quadratic  contributions  $\lambda_m\langle\sigma\rangle^2$ 
and  corrections suppressed by $1/\langle\sigma\rangle^2$, in addition to 
log-like terms $\ln\langle\sigma\rangle\sim\ln\mu$ present in the ``usual'' DR scheme. 
Our method to generate the subtraction scale via spontaneous breaking
in a dilaton-modified DR can also be implemented in other regularizations.
 Also note  that the role of $\mu\sim \sigma$ 
as a finite, non-zero ``DR scale'' means that only non-zero, finite 
 $\langle\sigma\rangle$ is allowed.  In fixing its actual numerical value,  Planck scale  physics
(gravity) is expected to play a role.

Although we do not explore them here, our results can have 
interesting  applications to phenomenology, such as model 
building beyond Standard Model (SM) \cite{AF,EG,GGR}.
For reference only, we provide below
the one-loop potential in the scale invariant version of the SM\footnote{This is just the SM with 
no classical mass term for the higgs in the Lagrangian.} extended 
by the dilaton. With the usual Coleman-Weinberg (CW) part $\delta U_\textsf{CW}$\footnote{
$\delta U_\textsf{CW}= 
\frac{1}{64\pi^2}
\sum_i  N_i M^4_i \big[\ln M^2_i/\mu^2(\phi,\sigma)-C_i\big]$,
$i=(G,S,W,Z,t)$  for Goldstone bosons, real scalars, gauge bosons, top, 
respectively, with $(N_G,N_S,N_W,N_Z,N_t)=(3,1,6,3,-12)$. 
 $C_i=3/2$ for fermions or scalars and $5/6$ for gauge bosons.
$M^2_G= \lambda_\phi\phi^2+\lambda_m \sigma^2$,
$M^2_W= \frac{1}{4} g^2 \phi^2$,
$M^2_Z=\frac{1}{4} (g^2+g^{\prime 2}) \phi^2$,
$M^2_t= \frac{1}{2}  y^2_t \phi^2$.
The potential (\ref{v}) of Higgs-dilaton:
$V\!=\!\lambda_\phi\vert H\vert^4\!+\!\lambda_m\vert H\vert^2\sigma^2\!+\!
 (\lambda_\sigma/4)\sigma^4$
with $H\!=\!(0,\phi)/\sqrt 2$ (unitary gauge).},
 the one-loop scalar potential in the SM is (with $M_\phi^2$, $M_\sigma^2$
as in eq.(\ref{ms}))
\medskip
\bea
 U&=&
 \frac{\laf}{4}\,\phi^4+\frac{\lam}{2}\phi^2\sigma^2+\frac{\las}{4}\sigma^4
+\delta U_\textsf{CW}\nonumber\\
&+&\!\!\!
\frac{1}{64\pi^2} \,\Big\{\,
\laf\lam \frac{\phi^6}{\sigma^2}-\big(16 \laf\lam+6\lam^2-3 \laf\las\big)\phi^4
-\big(16\lam+25\las\big)\,\lambda_m\,\phi^2\sigma^2- 21\las^2\sigma^4
\nonumber\\[4pt]
&+&\!\!\!
\!\! M_\sigma^4\ln \frac{M_\sigma^2}{z^2\sigma^2}
- \frac{3}{2} \,\Big[ ( 9 \laf^2\!+\lam^2)\phi^4+2\lam \,(3\laf\!+4\lam\! +3\las) \phi^2\sigma^2
+(\lam^2\!+\!9\las^2)\,\sigma^4\Big]\!
\Big\}
\eea

\medskip\noindent
The last two lines give the new correction $\Delta U$,
while in all CW terms one uses $\mu\!=\!z\sigma$ with
$z\!=\!\langle\phi\rangle/\langle\sigma\rangle$.
This equation can be used as the starting point in  phenomenological studies.

\section{Conclusions}\label{s7}

Scale invariant theories 
are  often considered to address the hierarchy problem of the 
Standard Model. However,  the regularization  of their quantum corrections  
breaks   explicitly the scale symmetry that one wants to study.  This is because all
 regularizations  introduce  a dimensionful parameter e.g. the  couplings in DR,
the UV scale in other regularizations, etc.  One can avoid this problem by using 
a manifestly scale invariant regularization in which the Goldstone mode of this symmetry
(dilaton) plays a central role. We used a dilaton-dependent subtraction 
 {\it function} $\mu=\mu(\sigma)$   that replaces  the ordinary subtraction {\it scale}. 
We applied this procedure to the  DR scheme,  to obtain
a scale invariant   one-loop scalar potential $U(\phi,\sigma)$
for the (higgs-like)  scalar $\phi$ and dilaton $\sigma$.
After  spontaneous breaking of  scale invariance    when $\langle\sigma\rangle\not=0$, 
all mass scales of the theory, including the ``usual'' subtraction scale, 
are generated from  this single vev.

The scale invariance of the action in $d=4-2\epsilon$ 
and the usual rescaling $\lambda\ra\mu^{2\epsilon} \lambda$ that ensures
dimensionless quartic couplings,  change the potential
$V(\phi,\sigma)\ra\mu(\sigma)^{2\epsilon} V(\phi,\sigma)$ which  now  contains new  
(``evanescent'') interactions due to the field dependence of $\mu$.  
At the quantum level, these  interactions generate  new,  finite corrections.

We found a new, (finite) one-loop correction $\Delta U$ 
to the potential, overlooked by previous studies, that is present
 beyond the usual Coleman-Weinberg term which is also
modified into a scale invariant form. 
For the minimal case  $\mu(\sigma)=z \,\sigma$, 
it was shown that  $\Delta U$ also  contains a 
non-polynomial operator  $\propto\phi^6/\sigma^2$
 with a known, finite coupling. After spontaneous breaking of
scale invariance, this operator generates a series of (polynomial) terms
suppressed by powers of $\langle\sigma\rangle\not=0$. 
At higher loop orders, more such  operators are expected.

Technically,   $\Delta U$ is generated from an  evanescent correction ($\propto\!\epsilon$) 
to the field-dependent masses of the states  ``running'' in the loop
correction to the potential,  which cancels 
the pole ($\propto\! 1/\epsilon$) of the momentum integral, to
give rise to a finite $\Delta U$. And since it is finite, $\Delta U$
was found to be independent of the subtraction   (dimensionless) parameter ($z$).
Of course {\it physics} must be independent of the parameter $z$
and of  the subtraction scale $\mu(\langle\sigma\rangle)=z\langle\sigma\rangle$ 
after spontaneous breaking  of scale symmetry. We  showed this by verifying 
that  the  one-loop potential $U(\phi,\sigma)$  respects  the Callan-Symanzik equation.

 Further, the  correction from $\Delta U$ to the 
mass of the  higgs-like scalar $\phi$ remains under control (small) without additional
 tuning  beyond that  done at the  tree-level to enforce the hierarchy 
$\langle\phi\rangle\ll\langle\sigma\rangle$. It is possible that this 
 behaviour survives  in higher  orders, in a manifest scale invariant calculation.
This could provide a solution to the hierarchy problem beyond the one-loop order discussed here.

More general subtraction  functions that  depend on both $\sigma$ and $\phi$    
were ruled out since in  this case there are quantum operators that force  the  visible 
  ($\phi$) and hidden ($\sigma$) sectors
to interact in $d\!=\!4$ even in their classical decoupling limit!
Avoiding  this behaviour dictates that the subtraction scale is generated by the 
dilaton only ($\mu\sim \sigma$), as  considered above.

This study and the scale-invariant regularization are of interest   to 
 theories that study scale invariance at the quantum level.

\vspace{1cm}
\noindent
{\bf Acknowledgements:  }
The author thanks Hyun Min Lee,  E. Dudas and G. G. Ross for discussions on this topic.
This work  was supported by a grant of the Romanian National  Authority for
Scientific Research (CNCS-UEFISCDI) under  project number PN-II-ID-PCE-2011-3-0607.

\end{document}